\documentclass{amsproc}

\usepackage{amssymb}

\usepackage{graphicx}

\usepackage{amsmath,amssymb,amsfonts,amsxtra,xfrac,footnote,rotating, lscape, pgfplots,tikz}
\usepackage[inline]{enumitem}

\usepackage{graphicx}
\usepackage{caption}
\usepackage{subcaption}

\usetikzlibrary{positioning}
\usetikzlibrary{chains}
\usetikzlibrary{arrows,fit,decorations.pathreplacing}
\tikzstyle{every picture}+=[remember picture]
\tikzstyle{na} = [baseline=-.5ex]

\usetikzlibrary{arrows, fit, decorations.markings, calc, fadings, decorations.pathreplacing, patterns, decorations.pathmorphing, positioning}

\newtheorem{theorem}{Theorem}[section]

\theoremstyle{definition}

\theoremstyle{remark}

\numberwithin{equation}{section}

\def\tilde{\widetilde}
\def\t{\tilde}
\def\hat{\widehat}
\def\h{\hat}
\def\bar{\overline}

\def\rt2{\sqrt{2}}

\def\d{\partial}

\def\det{\mathop{\rm det}}

\def\Tr{\mathop{\rm Tr}}

\def\rk{\mathrm{rk}}
\def\dim{\mathrm{dim}}

\def\cF{\mathcal{F}}
\def\cH{\mathcal{H}}
\def\cM{\mathcal{M}}
\def\cN{\mathcal{N}}
\def\cQ{\mathcal{Q}}
\def\cO{\mathcal{O}}
\def\cR{\mathcal{R}}
\def\cV{\mathcal{V}}
\def\cW{\mathcal{W}}
\def\cG{\mathcal{G}}

\def\bC{\mathbb{C}}
\def\bR{\mathbb{R}}
\def\bZ{\mathbb{Z}}
\def\bP{\mathbb{P}}

\newcommand{\be}{\begin{equation}} \newcommand{\ee}{\end{equation}}
\newcommand{\bea}{\begin{equation} \begin{aligned}} \newcommand{\eea}{\end{aligned} \end{equation}}

\newcommand{\PE}{\mathop{\rm PE}}
\newcommand{\PL}{\mathop{\rm PL}}

\begin{document}
	
	\title{3d supersymmetric gauge theories and Hilbert series}
	
	
	\author{Stefano Cremonesi}
	\address{Department of Mathematics, King's College London,
		The Strand, London WC2R 2LS, United Kingdom}
	\address{Department of Mathematical Sciences, Durham University, Durham DH1 3LE, United Kingdom}
	\email{stefano.cremonesi@durham.ac.uk}
	\thanks{To appear in the Proceedings of String-Math 2016. } 
	
	
	\date{\today}
	
	\begin{abstract}
		The Hilbert series is a generating function that enumerates gauge invariant chiral operators of a supersymmetric field theory with four supercharges and an $R$-symmetry. In this article I review how  counting dressed 't Hooft monopole operators leads to a formula for the Hilbert series of a 3d $\cN\geq 2$ gauge theory, which captures precious information about the chiral ring and the moduli space of supersymmetric vacua of the theory. (Conference paper based on a talk given at String-Math 2016, Coll\`ege de France, Paris.)
	\end{abstract}
	
	\maketitle
	
	
	\section{Introduction}\label{sec:intro}
	
	There is a long and illustrious tradition of fruitful interplay between supersymmetric quantum field theory and geometry \cite{Hitchin:1986ea,IntriligatorSeiberg1996,LutyTaylor1996,IntriligatorSeiberg1996}. The main bridge between the two topics is the concept of the moduli space of supersymmetric vacua, the set of zero energy configurations of the field theory, which in the context of supersymmetric field theories with at least four supercharges is a complex algebraic variety equipped with a K\"ahler metric. 
	
	Moduli spaces of vacua of quantum field theories with four supercharges in four spacetime dimensions have been studied in great detail since the 1990's, and their algebro-geometric structure is well understood. Less understood are their  counterparts in three dimensions, due to new scalar fields which are obtained by dualizing vectors in three dimensions. Interesting results on the moduli spaces of vacua of three-dimensional theories with four supercharges were obtained by means of semiclassical analysis \cite{BoerHoriOz1997,AharonyHananyIntriligatorEtAl1997}, but a precise understanding of the underlying algebraic geometry was lacking, except for a few simple theories \cite{Borokhov:2002cg}. 
	
	In this article I will review recent developments that allow one to make exact statements on the algebraic geometry of the moduli spaces of supersymmetric vacua of three-dimensional gauge theories with four or more supercharges \cite{Cremonesi:2013lqa,Cremonesi:2014kwa,Cremonesi:2014xha,Cremonesi2015a,CremonesiMekareeyaZaffaroni2016}. The key idea is to count the gauge invariant chiral operators that parametrize the moduli space of supersymmetric vacua, using a generating function called the Hilbert series. In the context of three-dimensional supersymmetric field theories, the gauge invariant chiral operators are dressed 't Hooft monopole operators: I will describe their properties and how to count them, leading to a formula for the Hilbert series. A peculiarity of 't Hooft monopole operators, that hindered the understanding of the algebraic geometry of moduli spaces of vacua of three-dimensional supersymmetric gauge theories, is that they obey relations that arise in the quantum theory. However, by applying plethystic techniques to the Hilbert series that counts dressed monopole operators, one can deduce information on the charges of generators and relations of the chiral ring of the field theory, which is the coordinate ring of its moduli space of vacua. The formalism is therefore capable of producing predictions on the quantum relations between monopole operators, without making any assumptions on them. 
	
	The article is organized as follows. In section \ref{sec:modulispace} I recall the concepts of moduli space of vacua and of chiral ring in the familiar ground of four-dimensional theories with four supercharges ($4d$ $\cN=1$ theories). In section \ref{sec:HS} I introduce the Hilbert series in that context and give a few useful examples for what follows. In section \ref{sec:3d4d} I introduce three-dimensional theories with four supercharges ($3d$ $\cN=2$ theories) and contrast them with $4d$ $\cN=1$ theories. In section \ref{sec:monopole} I discuss supersymmetric 't Hooft monopole operators and some of their properties. In section \ref{sec:HS_3d} I present the main formula \eqref{monopole_formula_N=2} for the Hilbert series, that counts dressed monopole operators. In section \ref{sec:CB} I apply these ideas to Coulomb branches of $3d$ $\cN=4$ theories, and in section \ref{sec:3d_Neq2_examples} to moduli spaces of vacua of $3d$ $\cN=2$ Yang-Mills and Chern-Simons theories. I conclude with a few remarks and open questions in section \ref{sec:conclusion}.

	\section{Moduli space of supersymmetric vacua and chiral ring}\label{sec:modulispace}
	
	Let us first recall some well-known facts about four-dimensional gauge theories with four Poincar\'e supercharges ($4d$ $\cN=1$ theories).
	Most of the structure of $4d$ $\cN=1$ supersymmetric theories carries over to $3d$ $\cN=2$ supersymmetric theories, which will be our main focus in the following.
	
	The $4d$ $\cN=1$ supersymmetry algebra consists of the following generators: the Lorentz generators $M_{\mu\nu}=-M_{\nu\mu}$ ($\mu,\nu=0,\dots,3$) for rotations and boosts, which generate the Lorentz group $SO(1,3)$; the momentum $P_\mu$, an $SO(1,3)$ vector which generates translations in the Minkowski spacetime $\bR^{1,3}$; the complex supercharges $\cQ_\alpha$ and $\bar{\cQ}^{\dot\alpha}=(\cQ_\alpha)^\dagger$ ($\alpha=1,2$ and $\dot\alpha=1,2$),  left-handed and right-handed Weyl spinors transforming in the two-dimensional representations $[1;0]$ and $[0;1]$ of the double cover of the Lorentz group $\mathrm{Spin}(1,3)=SL(2,\bC)\cong SL(2,\bR)\times SL(2,\bR)$, which generate translations along the Grassmann odd directions of superspace; and possibly $R$, a Lorentz scalar which generates a $U(1)_R$ symmetry that acts non-trivially on the supercharges.
	The commutation relations 
	are those required by the Lorentz properties of the generators recalled above, together with
	\be\label{4d_N=1_alg}
	\begin{split}
		\{\cQ_\alpha,\bar{\cQ}_{\dot{\alpha}}\}&= 2 (\sigma^\mu)_{\alpha\dot{\alpha}} P_\mu \\
		[R,\cQ_\alpha]=-\cQ_\alpha~,  & \qquad\quad~ [R,\bar{\cQ}_{\dot \alpha}]=\bar{\cQ}_{\dot\alpha}~,
	\end{split}
	\ee
	where $(\sigma^\mu)_{\alpha\dot{\alpha}}=(-\mathbf{1},\sigma_i)_{\alpha\dot{\alpha}}$ and $\sigma_i$ are the Pauli matrices that satisfy $\sigma_h \sigma_j = \delta_{hj}+i \epsilon_{hjk}\sigma_k$.
	In addition there might be a global non-$R$ symmetry algebra, often called flavour symmetry, generated by scalar charges which commute with the generators of the supersymmetry algebra.
	
	Fields in a $4d$ $\cN=1$ supersymmetric field theory fit in irreducible representations of the $4d$ $\cN=1$ superalgebra, which in turn constrains the form of interactions. Altogether, a $4d$ $\cN=1$ supersymmetric Lagrangian gauge theory is specified by the following data:  \begin{enumerate}
		\item {\bf Gauge group}: a compact semisimple Lie group $G$, to which one associates real vector multiplets $V^a$, with $a=1,\dots,\rk(G)$; 
		\item {\bf Matter content}: a representation $\cR$ of $G$, to which one associates complex chiral multiplets $X^i$, with $i=1,\dots,\dim(\cR)$;
		\item {\bf Superpotential}: a $G$-invariant holomorphic polynomial $W(X)$ in the matter fields.
	\end{enumerate}
	
	The scalar fields in the chiral multiplets, which we also denote as $X^i$ with a common abuse of notation, interact through a potential
	\be\label{scalar_potential}
	V= \sum_i |F_i|^2 + \frac{g^2}{2}\sum_a (D^a)^2~.
	\ee
	Here $F_i(X) = \frac{\d W}{\d X^i}$ are the $F$-terms, which equal the derivatives of the superpotential, $D^a(X,X^\dagger)=\sum_i X_i^\dagger (T^a)^i{}_j X^j$ are the $D$-terms, which equal the moment maps of the action of the gauge group $G$ on the matter representation $\cR$, and $g$ is the Yang-Mills coupling constant.%
	\footnote{In the $D^2$ term I have used a basis of the Lie algebra that diagonalizes the Killing form. If the gauge group $G$ is semisimple there is one Yang-Mills coupling constant per simple factor.}
	For $U(1)$ gauge factors, the moment map can be shifted by a constant, the Fayet-Iliopoulos (FI) parameter $\xi$, so that $D^{U(1)}=\sum_i q_i |X^i|^2 - \xi$, with $q_i$ the $U(1)$ charge of $X^i$.
	
	One can associate to a supersymmetric gauge theory specified by these data an object of great physical and mathematical interest: its moduli space of supersymmetric vacua $\cM$ \cite{LutyTaylor1996}. Physically, $\cM$ controls the low energy behaviour of the quantum field theory. Many of the impressive results on the dynamics of supersymmetric field theories obtained in the 1990's were indeed rooted in the analysis of their moduli spaces of vacua \cite{IntriligatorSeiberg1996}. Mathematically, $\cM$ provides a natural bridge between supersymmetric field theories and (differential and algebraic) geometry. Physically based results on moduli spaces of vacua of supersymmetric field theories can thus lead to interesting mathematical predictions. 
	
	Concretely, the {\bf moduli space of supersymmetric vacua} $\cM$ is defined as the set of constant field configurations that minimize the potential \eqref{scalar_potential}, modulo gauge equivalence:
	\bea\label{moduli_space}
	\cM &= \{ (X,X^\dagger) | ~F_i(X)=0 ~\forall i, ~ D^a(X,X^\dagger)=0 ~\forall a \}/G = \cF//G \\
	&\cong \{ (X) | ~\d W(X)=0 \}/G^\bC = \cF/G^\bC~.
	\eea
	$4d$ $\cN=1$ supersymmetry implies that the moduli space of vacua $\cM$ is a (possibly singular) K\"ahler manifold. %
	The first line of \eqref{moduli_space} expresses $\cM$ as a symplectic (in fact K\"ahler) quotient of $\cF$ by the gauge group $G$, whereas the second line expresses $\cM$ equivalently as a holomorphic (GIT) quotient by the complexified gauge group $G^\bC$. Here 
	\be\label{Fflat}
	\cF=\{ (X) | ~\d W(X)=0 \}
	\ee
	is the space of solutions of the $F$-term equations, often called $F$-flat moduli space or master space \cite{Forcella:2008bb}. Algebraically, it is a complex affine variety defined by the vanishing of the $F$-term relations $\d W(X)=0$.
	
	In the following I will adopt the holomorphic viewpoint in the second line of \eqref{moduli_space} and view the moduli space of vacua $\cM$ as a complex algebraic variety. $\cM$ is typically an affine variety. Of particular interest are superconformal field theories, whose moduli spaces of vacua are cones. The $\bC^*$ action whose radial part dilates the cone is the complexification of the $U(1)_R$ symmetry of the field theory.
	
	Closely related to the moduli space of supersymmetric vacua are the concepts of chiral operators and chiral ring \cite{LercheVafaWarner1989a}. Local gauge invariant {\bf chiral operators} $\cO_i(x)$ form a subset of observables in a $4d$ $\cN=1$ field theory  which are protected from quantum corrections. These are $\frac{1}{2}$-BPS operators that are annihilated by all the supercharges of positive $R$-charge: 
	\be\label{chiral_op}
	\bar\cQ_{\dot\alpha} \cO_i(x)=0 \qquad \forall~ \dot\alpha=1,2~.
	\ee 
	
	
	A crucial property of chiral operators is that their spacetime derivatives are $\bar\cQ$-exact and therefore vanish in expectation values, provided supersymmetry is unbroken. It follows that a product of chiral operators is free of short distance divergences and that its expectation value factorizes into the product of spacetime constant one-point functions: $\langle \cO_{i_1}(x_1)\dots \cO_{i_n}(x_n)\rangle=\langle \cO_{i_1}\rangle\dots \langle \cO_{i_n}\rangle $. 
	
	Chiral operators form a commutative ring, the {\bf chiral ring} $\cR$, with product  
	\be\label{chiral_ring}
	\cO_i \cO_j = c_{ij}{}^k \cO_k + \bar\cQ_{\dot \alpha}(\dots)^{\dot\alpha}~,
	\ee
	where the only spacetime dependence is in the $\bar\cQ$-exact term, and repeated indices are summed over. Since we are physically interested in taking expectation values, we will work at the level of $\bar\cQ$-cohomology and omit  $\bar\cQ$-exact terms in \eqref{chiral_ring} from now on. The chiral ring is then specified once a basis of chiral operators $\{\cO_i\}$ and the structure constants $c_{ij}{}^k$ are provided.
	
	The expectation values $\langle \cO_i \rangle$  of gauge invariant chiral operators, or equivalently $\bar\cQ$-cohomology classes, are holomorphic functions on the moduli space of vacua $\cM$. It is generally expected, though not proven to the best of my knowledge, that the correspondence between expectation values of chiral operators and holomorphic functions on $\cM$ is one-to-one, once relations are taken into account. With this physically motivated assumption, the chiral ring $\cR$ of the supersymmetric field theory is identified with the coordinate ring of its moduli space of vacua $\cM$. We would then like to characterize the chiral ring as a quotient ring
	\be\label{quotient_ring}
	\cR = \bC[\cO_1,\dots,\cO_n]/I~,
	\ee
	determine the generators $\cO_1,\dots,\cO_n$ of the polynomial ring and the defining relations of the ideal $I$. 
	
	In a 4d $\cN=1$ gauge theory, the chiral operators are $G$-invariant polynomials in the matter fields $X$.%
	\footnote{We neglect here glueball operators \cite{Cachazo:2002ry}, since they do not play a role in three dimensions.}
	If there is no gauge symmetry, the chiral ring is just the Jacobian ring of the superpotential $W$. In a gauge theory, however, the quotient by the gauge group in \eqref{moduli_space} makes it often hard to explicitly determine generators and relations of the chiral ring, and therefore the defining equations of the moduli space $\cM$ as an algebraic variety.

	\section{The Hilbert series}\label{sec:HS}
	
	Since determining generators and relations of the chiral ring of a supersymmetric gauge theory is in general a difficult task, it helps to exploit as much as possible the symmetries of the theory. 
	A very useful tool in this respect is 
	the {\bf Hilbert series} \cite{Benvenuti:2006qr},%
	\footnote{See also \cite{Pouliot1999} for an early incarnation of this concept.}
	a generating function that counts scalar gauge invariant chiral operators, graded by their charges under a Cartan subalgebra of the global symmetry:
	\be\label{HS}
	H(t,\hat x) = \mathrm{Tr}_\cH \bigg(t^R \prod_{\h i} \h x_{\h i}^{\h Q_{\h i}} \bigg)~.
	\ee
	Here $\cH=\{\cO_i | \bar\cQ_{\dot{\alpha}} \cO_i = 0, ~ M_{\mu\nu} \cO_i=0 \}$ denotes the vector space of gauge invariant scalar chiral operators, that parametrize the moduli space of supersymmetric vacua.%
	\footnote{An alternative count of protected operators is provided by the 
		``superconformal'' index \cite{Romelsberger2006}, which also counts fermionic and short non-chiral operators, and depends on the superpotential only through the $R$-charges of matter fields. Since our interest is in the moduli space of vacua and the chiral ring, we focus on the Hilbert series rather than the superconformal index. } 
	It can be decomposed into common eigenspaces of the $U(1)_R$ generator $R$ 
	and the generators $\h Q_{\h i}$ of the Cartan subalgebra of the flavour symmetry. For a superconformal field theory the $R$-charges can be taken positive, thus \eqref{HS} is a Taylor series in $t$ (and a Laurent series in $\h x_{\h i}$), and the eigenspaces are finite dimensional. The Hilbert series is a character on the vector space of scalar chiral operators: the coefficients of the series are the dimensions of the common eigenspaces of the global symmetry.
	
	In light of the correspondence between the space $\cH$  of scalar chiral operators of the supersymmetric field theory in \eqref{HS} and the space $\cH^0(\cM)$  of holomorphic functions on its moduli space of vacua $\cM$, the Hilbert series \eqref{HS} can be interpreted geometrically as a character of the action of the global symmetry group on $\cH^0(\cM)$.  
	For a superconformal field theory, whose moduli space of vacua is a cone, the Hilbert series equals the equivariant index of the Dolbeault operator on $\cM$ 
	\be\label{index-char}
	H(t,\hat x) = \mathrm{Tr} \bigg(t^R \prod_{\h i} \h x_{\h i}^{\h Q_{\h i}} ~\Big|~ \cH^0(\cM)\bigg) = \sum_p (-1)^p ~\mathrm{Tr} \bigg(t^R \prod_{\h i} \h x_{\h i}^{\h Q_{\h i}} ~\Big|~ \cH^p(\cM)\bigg)~,
	\ee
	dubbed index-character in \cite{Martelli:2006yb}, because higher Dolbeault cohomology groups $\cH^p(\cM)$ vanish for $p>1$. 
	In \eqref{HS}-\eqref{index-char} we have distinguished the $R$-charge, which generates a $\bC^\ast$ action that rescales the holomorphic top form of $\cM$, from the flavour charges $\h Q_{\h i}$, which  generate a torus action that leave the holomorphic top form invariant.
	
	Useful information on the moduli space of vacua $\cM$ can be extracted from the Hilbert series \cite{Benvenuti:2006qr}. For instance, the complex dimension $d$ of $\cM$ is the order of the pole at $t=1$ of the unrefined Hilbert series $H(t,1)$, and the coefficient of $(1-t)^{-d}$ is proportional to the volume of the $(d-1)$-dimensional base of $\cM$. Most importantly, the charges of the generators and relations can be extracted using plethystic techniques, if higher syzygies can be disentangled.%
	\footnote{This is often possible with some physical input, such as an independent determination of the dimension of the moduli space, and the help of computer algebra such as \texttt{Macaulay2} \cite{M2}.}
	Once this is achieved, the problem of presenting the moduli space $\cM$ as an algebraic variety (or equivalently the chiral ring as a quotient ring \eqref{quotient_ring}) is reduced to determining a finite number of coefficients, that specify which linear combinations of the  chiral operators having the appropriate charges appear as generators or relations.
	
	After this general discussion, let us see how the Hilbert series is computed in practice for a sample of $4d$ $\cN=1$ supersymmetric quantum field theories. We start by considering theories with no gauge group, so that the moduli space \eqref{moduli_space} coincides with the $F$-flat space \eqref{Fflat}. For the theory of a free chiral multiplet $X$ of $R$-charge $R[X]=r$, the moduli space is the complex plane, the chiral ring is the polynomial ring in one complex variable $\bC[X]$ and the Hilbert series is simply the geometric series counting powers of $X$:%
	\footnote{$\PE$ is the plethystic exponential, the generating function of symmetric powers. For a multivariate function $f(x_1, \dots,x_n)$ that vanishes at the origin, 
		$$
		\PE[f(x_1,\dots,x_n)] = \exp\bigg(\sum_{p=1}^\infty \frac{1}{p} f(x_1^p,\dots,x_n^p) \bigg)~.
		$$ 
		This implies that $\PE[\sum_i a_i \prod_j x_j^{b_{ij}}]=\prod_i (1-\prod_j x_j^{b_{ij}})^{-a_i}$. 	
		The inverse of the plethystic exponential is the plethystic logarithm $\PL$. For a multivariate function $g$ that equals $1$ at the origin,	
		$$
		\PL[g(x_1,\dots,x_n)] = \sum_{k=1}^\infty \frac{\mu(k)}{k} \log g(x_1^k,\dots,x_n^k) ~,
		$$
		where $\mu(k)$ is the M\"obius function.
	}
	\be\label{free}
	H= 1 + \tau + \tau^2 + \dots = \frac{1}{1-\tau} = \PE[\tau] ~, \qquad \tau=t^{r}~.
	\ee
	
	If $X$ is subject to a superpotential $W(X)=X^{N+1}$, 
	the Hilbert series becomes
	\be\label{X^{N+1}}
	H= 1 + \tau + \tau^2 + \dots+\tau^{N-1} = \frac{1-\tau^N}{1-\tau} = \PE[\tau-\tau^N] ~, \qquad \tau=t^{\frac{2}{N+1}}~.
	\ee
	The chiral ring is $\bC[X]/\langle X^N \rangle$ and the moduli space $\cM$ consists of a point of multiplicity $N$. The generator $X$ and the relation $X^N=0$ are respectively associated to the positive term $+\tau$ and the negative term  $-\tau^N$ in the argument of the plethystic exponential. In general, the plethystic logarithm of the Hilbert series terminates for theories whose moduli spaces are complete intersections, the dimension of which is the number of generators minus the number of relations. 
	
	The case of a generic polynomial superpotential $W(X)=X^{N+1} + \sum_{i=1}^N c_i X^{N-i}$ of degree $N+1$ can be treated similarly: even though the $U(1)_R$ symmetry is explicitly broken by the subleading terms in the superpotential, it can be restored by assigning $R$-charges to the parameters $c_i$, as is common practice in the analysis of supersymmetric field theories \cite{Seiberg1993}. The parameters $c_i$ are not dynamical and are not counted by the Hilbert series, which is insensitive to them and remains \eqref{X^{N+1}}, but they may (and do) appear in the relations. The Hilbert series only constrains the charges of the relations, which in this case must be of the form $X^N + \sum_i \alpha_i c_i X^{N-i}=0$, but does not fix the coefficients, which in this case we know to be $\alpha_i = (N-i)/(N+1)$. Of course there is no need to invoke the Hilbert series to study the chiral ring of such a simple theory, but this example makes it clear which information can be extracted from the Hilbert series (\emph{i.e.} charges of operators, generators and relations) and which cannot (\emph{i.e.} the precise form of the relations, unless they are entirely fixed by symmetry). Even when there are coefficients in the relations that cannot be determined by symmetry alone, the Hilbert series is a very useful tool for deducing the most general form of the chiral ring relations that is consistent with symmetry.

	Another simple but more interesting example of moduli space is provided by the XYZ model, a theory of three chiral multiplets $X$, $Y$ and $Z$ with the trilinear superpotential $W=XYZ$. From the $F$-term relations $\d W=0$ we deduce that the chiral ring is $\bC[X,Y,Z]/\langle YZ, ZX, XY\rangle$. The moduli space $\cM$ consists of three 1-dimensional components (in physical jargon ``branches'') parametrized by $X$, $Y$ and $Z$ respectively, meeting at a point. The Hilbert series reads
	\be\label{XYZ}
	\begin{split}
		H&=\frac{1}{1-\tau x}+ \frac{1}{1-\tau y}+ \frac{1}{1-\tau z}-2= \qquad\qquad (\tau=t^{2/3}~,\quad xyz=1)\\
		&= \frac{1-\tau^2(yz+xz+xy)+2\tau^3 xyz}{(1-\tau x)(1-\tau y)(1-\tau z)} ~.\\
		\PL[H]&= (x+y+z)\tau-(yz+xz+xy)\tau^2+2\tau^3-(x+y+z)\tau^4+\dots~.
	\end{split}
	\ee
	The first line shows that the moduli space $\cM$ consists of three copies of $\bC$ meeting at a point; the second line shows the three generators in the denominator, and that the moduli space is not a complete intersection because the numerator does not factorize. The plethystic logarithm in the last line allows us to extract information about the ring: the generators $X,Y,Z$ of the polynomial ring appear at order $\tau$, the generators of the ideal of relations by which we quotient $\d_X W,\d_Y W,\d_Z W$ at order $\tau^2$, and then we see higher order syzygies: $X\d_X W- Y\d_Y W$ and $X\d_X W- Z\d_Z W $ at order $\tau^3$, $\d_X W \d_Y W - Z^2 \d_Z W$ and cyclic permutations at order $\tau^4$, and so on. The plethystic logarithm is a series which does not terminate: this is the general structure for theories with non-complete intersection moduli spaces.
	
	Next we consider gauge theories. {\bf Gauging} a subgroup $G$ of the flavour symmetry leads in the holomorphic description of the moduli space in \eqref{moduli_space} to the quotient $\cM=\cF/G^\bC$ of the $F$-flat moduli space by the complexified gauge group. 
	At the level of the Hilbert series, the projection to gauge singlets is achieved by averaging the Hilbert series $H_\cF$ of the ungauged theory (whose moduli space is the $F$-flat moduli space $\cF$) over the gauge group
	\be\label{gauging}
	H (t,\h x)= \oint d\mu_G(x) H_\cF(t,x,\h x)
	\ee
	using the Haar measure 
	\be\label{Haar}
	\oint d\mu_G(x)= 
	\bigg(\prod_{j=1}^r \oint \frac{d x_j}{2\pi i x_j}\bigg) \prod_{\alpha \in \Delta_+} \left( 1- x^\alpha\right)~.
	\ee
	The integral is over the maximal torus of $G$, $r=\rk(G)$ is the rank of the gauge group, $\Delta_+$ is the set of positive roots of its Lie algebra, and I have used the short-hand $x^\alpha=\prod_{i=1}^r x_i^{\alpha_i}$. Gauge fugacities are denoted by $x$ and ungauged flavour fugacities by $\h x$.
	
	A simple class of examples, that will be useful in the following, is provided by theories with gauge group $G$ and a chiral multiplet $\Phi$ in the adjoint representation. (This is also the vector multiplet sector of $4d$ $\cN=2$ theories with gauge group $G$. The branch of the moduli space of vacua where the vector multiplet scalar $\Phi$ takes expectation values is called the Coulomb branch.) Setting $\tau=t^{R[\Phi]}$, the Hilbert series reads 
	\be\label{HS_adjoint}
	H(\tau)= \oint d\mu_G(x) \PE[\tau \chi_{ad}^G (x)] = \prod_{i=1}^r \frac{1}{1-\tau^{d_i(G)}} ~,
	\ee
	where $\chi_{ad}^G (x)$ is the character of the adjoint representation of the gauge group. The result expresses the well-known fact that the ring of invariants of the adjoint representation is freely generated by Casimir invariants $u_i$ of degrees $d_i(G)$. Hence $\cR=\bC[g]^G=\bC[\phi_1,\dots,\phi_r]/\cW_G=\bC[u_1,\dots,u_r]$. \emph{E.g.} for $G=SU(N)$ the Casimir invariants are $d_i(SU(N))=2, 3, \dots,N$. 
	
	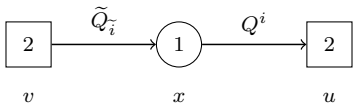
\begin{figure}
		\begin{tikzpicture}[baseline, baseline,font=\footnotesize, scale=1]
		\begin{scope}[auto,%
		every node/.style={draw, minimum size=.6 cm}, node distance=1.5cm];
		\node[rectangle] (FQt) at (-2,0) {$2$};
		\node[circle] (U1) at (0, 0) {$1$};
		\node[rectangle] (FQ) at (2,0) {$2$};
		\end{scope}
		\draw[draw=black,solid,line width=0.2mm,->]  (U1) to node[midway,above] {$Q^{i}$}node[midway,above] {}  (FQ) ; 
		\draw[draw=black,solid,line width=0.2mm,->]  (FQt) to node[midway,above] {$\t Q_{\t i}$}node[midway,above] {}  (U1) ;
		
		\node at (-2,-0.7) {$v$};
		\node at (0,-0.7) {$x$};
		\node at (2,-0.7) {$u$};
		
		\end{tikzpicture}
		\caption{Quiver diagram for SQED with two flavours. \label{SQED_2}}
	\end{figure}

	Another example is a $U(1)$ gauge theory with two matter fields $Q^i$ of charge $1$ and two matter fields $\t Q_{\t i}$ of charge $-1$, also known as SQED with two flavours. See figure \ref{SQED_2} for the quiver diagram of this theory. It turns out to be interesting to compute the Hilbert series in the presence of a background electric charge $-B$ for the $U(1)$ gauge symmetry, even though this may seem artificial from the perspective of four-dimensional gauge theory. (We will see that $B$ has a more natural interpretation in three dimensions.) 
	This modified Hilbert series, often called baryonic Hilbert series \cite{Forcella:2007wk}, counts polynomials in the matter fields $Q$ and $\t Q$ of total electric charge $B$ to compensate the background electric charge $-B$, and is computed by the formula \cite{Forcella:2007wk}
	\be\label{resolved_conifold}
	\begin{split}
		H_{-B}(\tau,u,v)&\equiv g_1(\tau,u,v;B) = \oint \frac{dx}{2\pi i x} x^{-B} \PE\left[\tau x\Big(u+\frac{1}{u}\Big)+\tau \frac{1}{x}\Big(v+\frac{1}{v}\Big)\right] \\
		&=\begin{cases}
			\sum_{n=0}^\infty [n+B;n]_{u,v} \tau^{2n+B} & B\geq 0\\
			\sum_{n=0}^\infty [n;n-B]_{u,v} \tau^{2n-B} & B\leq 0
		\end{cases}~.
	\end{split}
	\ee
	Here $u$ and $v$ are fugacities for the $SU(2)_u\times SU(2)_v$ flavour symmetries that rotate $Q^i$ and $\t Q_{\t i}$ respectively, and $[n;m]_{u,v}$ denotes the character of the representation $[n;m]$ of $SU(2)_u\times SU(2)_v$.

	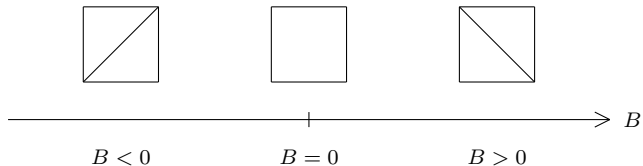
\begin{figure}
		\begin{tikzpicture}[baseline, baseline,font=\footnotesize, scale=1, transform shape]
		\draw (-3,1.5)--(-2,1.5); 
		\draw (-3,.5)--(-2,.5);
		\draw (-3,1.5)--(-3,.5); 
		\draw (-2,1.5)--(-2,.5);
		\draw (-2,1.5)--(-3,.5);
		
		\draw (-.5,1.5)--(.5,1.5); 
		\draw (-.5,.5)--(.5,.5);
		\draw (-.5,1.5)--(-.5,.5); 
		\draw (.5,1.5)--(.5,.5);
		
		\draw (3,1.5)--(2,1.5); 
		\draw (3,.5)--(2,.5);
		\draw (3,1.5)--(3,.5); 
		\draw (2,1.5)--(2,.5);
		\draw (2,1.5)--(3,.5);
		
		\draw (-4,0)--(4,0);
		\draw (3.8,.1)--(4,0);
		\draw (3.8,-.1)--(4,0);	
		\draw (0,.1)--(0,-.1);
		
		\node at (4.3,0) {$B$};
		
		\node at (-2.5,-0.5) {$B<0$};
		\node at (0,-0.5) {$B=0$};
		\node at (2.5,-0.5) {$B>0$};
		
		\end{tikzpicture}
		\caption{Line bundles on the conifold and resolutions. \label{fig_resol_conif}	}	
	\end{figure}
	

	The background electric charge or ``baryonic charge'' $B$ in \eqref{resolved_conifold} is a discrete analogue of the Fayet-Iliopoulos parameter $\xi$ introduced after \eqref{scalar_potential}, which leads to a resolution of the conical moduli space of vacua in the symplectic reduction in the first line of \eqref{moduli_space}. The theory that we are discussing is nothing but the gauged linear sigma model for the conifold, albeit viewed as a four-dimensional theory: its moduli space of vacua $\cM$ is the conifold if $\xi=0$, and the resolved conifold if $\xi\neq 0$, with the resolutions at $\xi>0$ and $\xi<0$ being related by a flop transition. In the holomorphic language, the Hilbert series \eqref{resolved_conifold} with insertion of the baryonic charge $B$ counts holomorphic sections of the line bundle $\cO(B D)$, where $D$ is the toric divisor associated to $Q$ fields, and $-D$ the toric divisor associated to $\t Q$ fields. See figure \ref{fig_resol_conif} for a summary.
	The baryonic Hilbert series counts operators of the schematic form  $Q^B (Q \tilde Q)^n$ for $B\geq0$ and $\t Q^{-B} (Q \tilde Q)^n$ for $B\leq 0$. 
	For $B=0$ we obtain the Hilbert series of the conifold
	\be\label{conifold}
	H_0(\tau,u,v)=\PE[\tau[1;1]_{u,v}-\tau^2]~,
	\ee 
	corresponding to the ring $\bC[M^1_1,M^1_2,M^2_1,M^2_2]/\langle M^1_1M^2_2-M^1_2 M^2_1\rangle$ generated by the four mesons $M^i_{\t i}=Q^i \t Q_{\t i}$ subject to a singlet relation $\det(M)=0$. We have thus recovered the algebraic description of the conifold \cite{CandelasOssa1990}. 
	
	
	
	\section{$3d$ $\cN=2$ gauge theories vs $4d$ $\cN=1$ gauge theories}\label{sec:3d4d}
	
	Moduli spaces of supersymmetric vacua of four-dimensional supersymmetric gauge theories, which have been described so far and analysed with the help of the Hilbert series, are of mathematical and physical interest and many nontrivial results have been obtained \cite{Benvenuti:2006qr,Feng:2007ur,Gray:2008yu,Hanany:2008kn,HananyMekareeyaTorri2010}. However, the construction is limited on both fronts. Mathematically, even though supersymmetry only requires the moduli space of vacua $\cM$ of $4d$ $\cN=1$ theories to be a K\"ahler manifold (or hyperkähler for the Higgs branch of 4d $\cN=2$ theories), the latter is actually a K\"ahler quotient (or hyperkähler quotient for the Higgs branch of 4d $\cN=2$ theories) for a Lagrangian supersymmetric gauge theory. On the physical side, the computation of the Hilbert series is essentially classical and reduces to counting gauge invariant polynomials in the matter fields appearing in the UV Lagrangian.%
	\footnote{Quantum corrected moduli spaces of vacua of several four-dimensional supersymmetric gauge theories were intensively studied in the 1990's \cite{IntriligatorSeiberg1996}, but the Hilbert series is to a large extent insensitive to the complex structure deformations induced by quantum effects.
	}
	
	Moduli space of vacua which are not (hyper)K\"ahler quotients occur for non-Lagrangian supersymmetric theories in four dimensions, and string theory or M-theory constructions have been used to study some of their properties, along the lines of \cite{Gaiotto:2009we}. We will pursue here an alternative way to overcome the limitations explained above, by considering three-dimensional supersymmetric gauge theories with at least four supercharges ($3d$ $\cN\geq 2$) instead of four-dimensional theories. 
	
	The $3d$ $\cN=2$ supersymmetry algebra can be obtained by dimensional reduction of the $4d$ $\cN=1$ supersymmetry algebra and is therefore very similar to \eqref{4d_N=1_alg}. The main modifications are that the supercharges are complex conjugate $3d$ Dirac spinors $\cQ_\alpha$ and $\bar{\cQ}_{\beta}$ transforming as doublets of $SL(2,\bR)$, and that the anticommutator of two supercharges is
	\be\label{3d_N=2_alg}
	\{\cQ_\alpha,\bar\cQ_\beta\} = 2 (\gamma^\mu)_{\alpha\beta} P_\mu + 2i \epsilon_{\alpha\beta} Z~. 
	\ee
	Here $\gamma^\mu = (-\mathbf{1},\sigma_1, \sigma_3)$, and $Z$ is a real generator of the centre of the supersymmetry algebra. The central charge $Z$ originates from the momentum along the reduced dimension. In the same vein, the vector and chiral multiplets of $3d$ $\cN=2$ supersymmetry can be obtained by dimensional reduction of the vector and chiral multiplets of $4d$ $\cN=1$ supersymmetry. Supersymmetric actions in three dimensions take the same form as in four dimensions, except for the possibility, peculiar to odd spacetime dimensions, to add supersymmetric Chern-Simons (CS) terms
	\be\label{CS_susy}
	S_{CS} = \frac{k_{ab}}{4\pi} \int d^3 x~ d^4 \theta ~\Sigma_a V_b = \frac{k_{ab}}{4\pi} \int (A_a d A_b + \sigma_a D_b +\mathrm{superpartners})~,
	\ee
	where $\Sigma_a = \epsilon^{\alpha\beta} \bar{D}_\alpha D_\beta V_a$ is the field strength multiplet  and $A_a$ is the gauge connection for a $U(1)$ gauge group. The real scalar $\sigma_a$ is the lowest component of the field strength  multiplet $\Sigma_a$, and originates from the component $A_4$ of the gauge connection in the dimensional reduction from four dimensions. 
	\eqref{CS_susy} is a Chern-Simons interaction between abelian gauge groups, but can be generalized to nonabelian gauge groups $G$ in a straightforward way using the Killing form as the symmetric pairing. 
	
	Despite the similarities between $4d$ $\cN=1$ and $3d$ $\cN=2$ supersymmetric gauge theories at the classical level, they behave very differently quantum-mechanically. 
	In what follows I will describe how the quantum physics of three-dimensional $\cN\geq 2$ supersymmetric gauge theories leads to moduli spaces of supersymmetric vacua $\cM$ which are (hyper)K\"ahler, as required by supersymmetry, though not (hyper)K\"ahler quotients.
	Unlike most of the previous results on moduli spaces of vacua of $3d$ supersymmetric gauge theories which were based on semiclassical analysis (see \cite{Intriligator:1996ex,deBoer:1996mp,BoerHoriOz1997,AharonyHananyIntriligatorEtAl1997,DoreyTong2000,Tong2000,Intriligator:2013lca} for a partial sample), the aim here will be to understand these moduli spaces as algebraic varieties and to develop general methods to characterize the chiral rings of such theories. In particular, I will review explicit formulae for the Hilbert series of the Coulomb branch of the moduli space of vacua of $3d$ $\cN=4$ gauge theories and for the Hilbert series of the moduli space of vacua of $3d$ $\cN=2$ gauge theories, which have been obtained in a series of recent works \cite{Cremonesi:2013lqa,Cremonesi2015a,CremonesiMekareeyaZaffaroni2016} (see also \cite{Hanany:2015via}).  
	
	The count of holomorphic functions in the Hilbert series encodes information about the moduli space of vacua and hints at a new construction of (hyper)K\"ahler moduli spaces which is alternative to the standard (hyper)K\"ahler quotient. In the context of $3d$ $\cN=4$ gauge theories, the Hilbert series formula of \cite{Cremonesi:2013lqa} has spurred recent activity both on the physical  \cite{Cremonesi:2014kwa,Cremonesi:2014uva,Cremonesi:2014xha,DelZottoHanany2015,Dey:2014tka,BullimoreDimofteGaiotto2015,Mekareeya2015,BullimoreDimofteGaiottoEtAl2016,HananyKalveks2016,HananySperling2016,BullimoreDimofteGaiottoEtAl2016a,CartaHayashi2016,HananySperling2016a} and on the mathematical front \cite{Nakajima2015a,Nakajima2015b,BravermanFinkelbergNakajima2016,BravermanFinkelbergNakajima2016a,NakajimaTakayama2016,KoderaNakajima2016}, leading eventually to a mathematical definition of the Coulomb branch and to several other interesting developments (see also Nakajima's and Bullimore's talk at this conference). 
	
	The novelty compared to four-dimensional theories is that three-dimensional supersymmetric gauge theories contain chiral 't Hooft monopole operators, a new class of gauge invariant chiral operators which are not polynomials in the matter fields. Monopole operators are subject to relations that arise quantum-mechanically and cannot be obtained by differentiating a superpotential. It is difficult to directly determine these relations for general theories, although impressive results have been obtained by direct computation for simple theories using conformal field theory techniques \cite{Borokhov:2002cg}. Nevertheless, we will see that physical arguments lead to a general group-theoretic formula for the Hilbert series that completely bypasses this problem. Once the Hilbert series is computed, exactly or in a Taylor expansion, we can learn about the quantum relations among monopole operators by means of plethystic methods, as was sketched in the four-dimensional examples described above.

	\section{'t Hooft monopole operators} \label{sec:monopole}
	
	Before we can explain how to compute the Hilbert series of $3d$ $\cN\geq 2$ supersymmetric gauge theories, we need to understand some of the properties of supersymmetric 't Hooft monopole operators. 
	
	In recent years it has been increasingly appreciated  that local operators in a quantum field theory need not be expressible as polynomials in the microscopic fields that are used to write down the Lagrangian \cite{Kapustin:2005py}. One also needs to include ``disorder'' (or defect) operators, which may be defined by prescribing appropriate singular boundary conditions in the path integral. While the definition of local disorder operators appears to put them on a different footing from standard local ``order'' operators, this difference is an artefact of our choice of field variables in the description of the quantum field theory. All local operators are on the same footing in the quantum theory. There are by now numerous examples of dualities (quantum equivalences between classically different field theories) that map standard order operators to disorder operators and vice versa, going from sine-Gordon -- massive Thirring duality \cite{Coleman1975,Mandelstam1975}, T-duality and mirror symmetry \cite{GiveonPorratiRabinovici1994,HoriVafa2000} in two spacetime dimensions to Intriligator-Seiberg mirror symmetry (supersymmetric particle-vortex duality) \cite{Intriligator:1996ex} and Aharony duality \cite{Aharony1997} in three dimensions.
	
	In the context of three-dimensional gauge theories, the relevant local disorder operators are 't Hooft monopole operators (monopole operators henceforth) \cite{'tHooft:1977hy}, which are introduced in the Euclidean formulation of the theory and may be obtained by dimensionally reducing 't Hooft loop operators in four dimensions. To insert a monopole operator $V_m(x)$ in a correlation function, one path integrates over gauge field configurations with a Dirac monopole singularity at the insertion point $x$. In a $3d$ $\cN=2$ gauge theory, the monopole operator can be supersymmetrized by imposing singular boundary conditions for all the bosonic fields in the $3d$ $\cN=2$ vector multiplet. Using spherical coordinates $(r,\theta,\varphi)$ centred at $x$, we define a $\frac{1}{2}$-BPS {\bf bare monopole operator} by imposing the following singular boundary conditions as $r\to 0$ \cite{Borokhov:2002cg,Borokhov:2003yu}:
	\be\label{monopole_op}
	\begin{split}
		A_\pm &\sim \frac{m}{2} (\pm 1 -\cos\theta) d\varphi\\
		\sigma &\sim \frac{m}{2r} ~.
	\end{split}
	\ee
	In the first line of \eqref{monopole_op}, $A_\pm$ is the gauge connection in the north/south patch of a trivialization of the $G$-bundle over the $S^2$ surrounding the insertion point $x$. The Dirac monopole singularity is given by an embedding $U(1)\hookrightarrow G$, specified by the {\bf magnetic charge} $m \in \mathfrak{h}/\cW$, a constant element of the Cartan subalgebra $\mathfrak{h}$ of the gauge Lie algebra $\mathfrak{g}$, defined modulo Weyl reflections. Well-definedness of the gauge bundle requires the Dirac quantization condition \cite{Englert:1976ng,Goddard:1976qe}
	\be\label{Dirac_quant}
	e^{2\pi i m} = \mathbf{1}_G \qquad \Longrightarrow \qquad m \in \Gamma_{G^\vee}/\cW~,
	\ee
	hence the magnetic charge $m$ is an element of the magnetic weight lattice, the weight lattice of the Langlands \cite{Langlands} or GNO \cite{Goddard:1976qe} dual group $G^\vee$ of the gauge group $G$, modulo Weyl reflections. $m$ can be viewed as the highest weight of an irreducible representation of the dual group $G^\vee$, or equivalently as specifying a cocharacter in $\mathrm{Hom}(U(1),G)$. 
	
	The boundary condition for the gauge connection in the first line of \eqref{monopole_op} defines a monopole operator which does not preserve any supersymmetry. By further imposing the boundary condition in the second line of \eqref{monopole_op} for the real scalar $\sigma$ in the $3d$ $\cN=2$ vector multiplet (coming from the component of the $4d$ gauge field in the reduced dimension), we define a $\frac{1}{2}$-BPS monopole operator that sits in a chiral multiplet, like all the matter fields in the theory. It is a crucial fact \cite{Borokhov:2002cg} that there exists a single half-BPS bare monopole operator for each choice of magnetic charge $m\in \Gamma_{G^\vee}/\cW$.
	
	The bare monopole operator $V_m$ defined by the boundary conditions \eqref{monopole_op} is made gauge invariant by averaging over the Weyl group, if there are no gauge Chern-Simons terms. It is called bare because, as will be explained below, monopole operators can also be dressed by matter fields before they are made gauge invariant. 
	
	Note that in this construction a vector multiplet, containing the bosonic fields $A$, $\sigma$ appearing in \eqref{monopole_op} and their supersymmetric partners, is traded for a tower of chiral multiplets $V_m$, the monopole operators, labelled by their magnetic charges $m$. As gauge invariant chiral operators, the monopole operators $V_m$ can take expectation value on the moduli space of vacua $\cM$ of the $3d$ $\cN=2$ gauge theory. In a semiclassical description, this corresponds to the fact that the real scalar $\sigma$ in the vector multiplet can take expectation value and contribute to the moduli space. At a generic point of the Coulomb branch, where the adjoint $\sigma$ takes expectation value breaking $G$ to its maximal torus, the Cartan components $\sigma_i$ of $\sigma$ are complexified by dual photons $\tau_i$ defined by $d\tau_i = \ast F_i$. For large expectation values monopole operators can be expressed as $V_m \simeq \exp\big(m\cdot(\frac{\sigma}{g^2}+ i \tau)\big)$ up to quantum corrections, but this semiclassical expression for monopole operators breaks down at loci of enhanced gauge symmetry \cite{BoerHoriOz1997,AharonyHananyIntriligatorEtAl1997}. We will use instead the path integral definition of monopole operators  \eqref{monopole_op}, which is more implicit but is valid even in strongly coupled regions of the moduli space of vacua with unbroken nonabelian gauge symmetry.
	
	For later purposes, it is useful to repeat the previous construction in the presence of {\bf background magnetic charges} for background vector multiplets associated to the global non-$R$ symmetries of the $3d$ $\cN=2$ gauge theory. These include flavour symmetries acting on the matter fields, as in $4d$ $\cN=1$ field theories, but also topological symmetries special to three dimensions that only act on monopole operators. Vector multiplets associated to gauge symmetries are fluctuating dynamical fields, whereas vector multiplets associated to global symmetries are frozen external parameters. 
	We label these generalized monopole operators as $V_{m;\h m, B}$, where $m$ is the dynamical magnetic charge for the gauge symmetry $G$, and $\h m$ and $B$ are fixed background magnetic charges for the global flavour and topological symmetries. Supersymmetry relates magnetic charges and real scalars in \eqref{monopole_op}, leading to the following correspondence between quantized charges that characterize monopole operators and continuous real scalar fields or masses:
	\be\label{mag_charges-real_scalars}
	\begin{tabular}{ l c c c c }
		Gauge: & \qquad $m$ & $\longleftrightarrow$ & $\sigma$ & \qquad real scalar \\
		Flavour:	& \qquad$\h m$ & $\longleftrightarrow$ & $\h\sigma$ & \qquad real mass \\
		Topological: &\qquad	$B$ & $\longleftrightarrow$ & $\xi$ & \qquad FI parameter
	\end{tabular}
	\ee
	
	We have encountered the ``baryonic charge'' $B$ earlier in the context of $4d$ $\cN=1$ gauge theories in \eqref{resolved_conifold}. There $B$ was introduced as a background electric charge for the gauge symmetry, the counterpart in the holomorphic quotient of the Fayet-Iliopoulos parameter $\xi$, whose effect was to resolve the conical moduli space of vacua in the symplectic quotient description. In the context of $3d$ $\cN=2$ gauge theories, the FI parameter $\xi$ can be interpreted as a background real scalar in the vector multiplet for the topological symmetry, and $B$ is the associated background magnetic charge.

	\subsection{Charges of monopole operators}\label{subsec:charges}
	
	We have defined a set of new chiral operators, the monopole operators $V_m$, which together with the standard gauge invariant polynomials in the matter fields parametrize the moduli space of supersymmetric vacua of a $3d$ $\cN\geq 2$ gauge theory. Since we are going to count monopole operators in the Hilbert series, we need to know how they are (electrically) charged under the global symmetries of the theory.
	
	The first symmetry under which monopole operators are charged is the topological symmetry that was mentioned above. For a theory with gauge group $G$, the topological symmetry group is $G_J=Z(G)$, the centre of the gauge group. The magnetic charge $m$ of a monopole operator is an element of the magnetic weight lattice of the gauge group $G$, $\Gamma_{G^\vee}$, up to Weyl symmetry. The topological charge of the monopole operator $J(m)$ is the magnetic charge modulo elements of the coroot lattice of $G$. For example, if $G=U(N)$, the magnetic charge is $m=\mathrm{diag}(m_1,\dots,m_N) \in \bZ^N/S_N$, the topological symmetry group is $G_J=U(1)$ and the topological charge is $J(m)=\Tr(m)= \sum_i m_i \in \bZ$.
	
	Monopole operators also carry electric charges under other global (as well as gauge) symmetries at the classical level, if the theory has Chern-Simons couplings \eqref{CS_susy}. In general, the classical charges of monopole operators under the Cartan generators are given by
	\be\label{charges_class}
	Q_A^{cl}(M)=-\sum_B k_{AB}M_B~,
	\ee
	where I labelled by $\{M_A\}=\{m_i,\h m_{\h i}, B_i, 0\}$ all the magnetic charges allowed by supersymmetry: $m_i$ for the Cartan subalgebra of the gauge symmetry, $\h m_{\h i}$ for the flavour symmetry, $B_i$ for the topological symmetry, and $0$ for the $R$-symmetry. Formula \eqref{charges_class} is just the statement that magnetic charges lead to electric charges in the presence of Chern-Simons interactions \eqref{CS_susy}. The charges of monopole operators under continuous abelian subgroups of the topological symmetry can be included in \eqref{charges_class} by means of appropriate mixed gauge-topological Chern-Simons couplings.
	
	The classical charges \eqref{charges_class} receive quantum corrections of the form \cite{Borokhov:2002cg,ImamuraYokoyama2011,Benini2011}
	\be\label{charges_quant}
	Q_A^{q}(M)=-\frac{1}{2}\sum_{\psi_a} Q_A[\psi_a]|m_a^{\rm eff}(M)|~,
	\ee
	where the sum runs over all fermionic fields $\psi_a$ in matter chiral multiplets and vector multiplets, of charge $Q_A[\psi_a]$ and  ``effective mass''%
	\footnote{We are slightly abusing terminology here. The effective real mass of a chiral multiplet, which equal its central charge $Z$ appearing in the supersymmetry algebra \eqref{3d_N=2_alg}, is actually \eqref{eff_mass} with the magnetic charges $M_A$ replaced by the real scalars or masses $\Sigma_A$, according to the correspondence \eqref{mag_charges-real_scalars}. In the background of a half-BPS monopole operator, magnetic charges and real scalars are related as in the second line of \eqref{monopole_op}, therefore the effective real mass is proportional to \eqref{eff_mass} and inversely proportional to the distance $r$ from the  point where the monopole operator is inserted.}
	\be\label{eff_mass}
	m_a^{\rm eff}(M) = \sum_A Q_A[\psi_a] M_A~.
	\ee
	
	Adding up the classical contribution \eqref{charges_class} and the quantum contribution \eqref{charges_quant}, the total electric charges of monopole operators are 
	\be\label{charges_tot}
	Q_A^{\rm eff}(M)=-\sum_B k_{AB}^{\rm eff}(M)M_B~,
	\ee
	where the {\bf effective Chern-Simons levels}
	\be\label{eff_CS}
	k_{AB}^{\rm eff}(M)= k_{AB}+\frac{1}{2}\sum_{\psi_a} Q_A[\psi_a]Q_B[\psi_a]~\mathrm{sign}(m_a^{\rm eff}(M))
	\ee
	must be integer for gauge invariance. This in turn constrains the values of the bare CS levels $k_{AB}$.
	
	\subsection{Dressed monopole operators}
	
	So far we have discussed bare monopole operators, chiral operators which are defined in terms of vector multiplets. The boundary condition \eqref{monopole_op} breaks the gauge group $G$ to a subgroup $G_m$, the stabilizer of the magnetic charge: $G_m\cdot m =0$. We call $G_m$ the {\bf residual gauge group}. (A similar discussion applies to the flavour symmetry group $\h G$ and the topological symmetry group $G_J$.) In the background of the monopole operator $V_m$, the vector multiplets for $G_m$, associated to roots $\alpha$ such that $\alpha(m)=0$, are massless. 
	The vector multiplets for $G/G_m$, associated to roots $\alpha$ such that $\alpha(m)\neq 0$, are massive by the Higgs mechanism. When integrated out, they correct the $R$-charge of $V_m$ according to the formula \eqref{charges_quant}, where all fermions in the vector multiplet have $R$-charge $1$. 
	
	Translating the previous discussion in a mathematical formula, the contribution of a dynamical vector multiplet to the Hilbert series of a $3d$ $\cN=2$ gauge theory is
	\be\label{monopole_contr_vector}
	\prod_{\alpha \in \Delta_+} t^{-|\alpha(m)|}(1-x^\alpha)^{\delta_{\alpha(m),0}}~.
	\ee
	The first factor accounts for the correction to the $R$-charge of the monopole operator $V_m$ due to the massive vector multiplets; the second factor is the contribution to the Hilbert series of the residual gauge group $G_m$, whose fugacities $x$ are eventually integrated over.

	A similar discussion applies to matter fields, which transform in the representation $(\cR,\h\cR,1)$ of $G\times \h G\times G_J$ with weights $(\rho,\h\rho,0)$. In the background of the monopole operator, the matter fields neutral under the $U(1)$ subgroup of $G\times \h G$ singled out by $m$ have vanishing ``effective mass''%
	\be\label{eff_mass_2}
	m_{\rho,\h\rho}^{\rm eff}(m,\h m)=\rho(m)+\h\rho(m)~.
	\ee
	They can take expectation value and dress the bare monopole operator without spoiling its supersymmetry \cite{Cremonesi:2013lqa,Cremonesi2015a,CremonesiMekareeyaZaffaroni2016}. 
	We call these massless matter fields {\bf residual matter fields}. On the other hand, the matter fields with nonvanishing \eqref{eff_mass_2} are massive, cannot take expectation value and are integrated out. Their only effect is to correct the charges of the bare monopole operator quantum-mechanically according to formula \eqref{charges_quant}. 
	
	In formulae, the contribution to the Hilbert series of a $3d$ $\cN=2$ gauge theory of a matter chiral multiplet $X$ of $R$-charge $r$ transforming in the representation $(\cR,\h\cR,1)$ of $G\times \h G\times G_J$  is
	\be\label{monopole_contr_matter}
	\prod_{\rho,\h\rho} (t^{r-1}x^\rho \h x^{\h\rho})^{-\frac{1}{2}|\rho(m)+\h\rho(\h m)|}\PE[\delta_{\rho(m)+\h\rho(\h m),0}~t^r x^\rho \h x^{\h\rho}]~,
	\ee
	where we have assumed that the matter field is not subject to $F$-term relations descending from a superpotential in order to simplify the formula (see \cite{Cremonesi2015a,CremonesiMekareeyaZaffaroni2016} for the generalization). 
	The first factor accounts for the quantum correction to the charges of the monopole operator $V_m$ due to the massive matter fields; the second factor is the contribution  of the  massless residual matter fields.
	
	In conclusion, we can dress a bare gauge-variant monopole operator by a polynomial in the residual matter fields to construct an operator that is invariant under the residual gauge group $G_m$. The resulting dressed monopole operator is then made $G$-invariant by averaging over the action of the Weyl group $\cW_G/\cW_{G_m}$.

	\section{Monopole formula for the Hilbert series of $3d$ $\cN\geq 2$ gauge theories}\label{sec:HS_3d}
	
	We have collected all the necessary ingredients to write down a formula for the Hilbert series of a $3d$ $\cN=2$ supersymmetric gauge theory. The Hilbert series counts the gauge invariant chiral operators of the theory, which are dressed monopole operators, possibly in the presence of background magnetic charges $(\h m, B)$ for the flavour and topological symmetries:
	\be\label{HS_3d_N=2}
	H(t,\h x,z; \h m, B) = \Tr{}_{\cH_{\h m,B}}\bigg(t^R z^J \prod_{\h i} \h x_{\h i}^{\h Q_{\h i}}\bigg)~.
	\ee 
	$\cH_{\h m,B}$ denotes the vector space of scalar chiral monopole operators of fixed background magnetic charges $(\h m,B)$ for the flavour and topological symmetries $\h G\times G_J$. When $\h m=B=0$, \eqref{HS_3d_N=2} is the standard Hilbert series that counts gauge invariant chiral operators that parametrize the moduli space $\cM_0$ of the superconformal field theory to which the gauge theory flows at low energy. The background magnetic charges $(\h m,B)$ correspond to turning on real masses for the flavour symmetry and Fayet-Iliopoulos parameters, that lead to a resolution of the moduli space $\cM_0$ of the SCFT. The generalized Hilbert series with background magnetic charges then counts holomorphic sections of line bundles rather than holomorphic functions on $\cM_0$, analogously to the baryonic Hilbert series \eqref{resolved_conifold}.

	The  dressing of monopole operators is summarized in the data of a {\bf residual gauge theory} $T_{m;\h m, B}$ of massless fields in the background of a monopole operator of magnetic charges $(m;\h m, B)$:
	\begin{enumerate}
		\item A residual gauge group $G_m$ (and flavour group $\h G_{\h m}$);
		\item Residual matter fields transforming in representations of $G_m\times \h G_{\h m}$;
		\item A residual superpotential $W_m$, which is obtained by setting to zero all the massive matter fields in the original superpotential $W$;
		\item Background electric charges $Q_i(m,\h m,B)$ under the Cartan generators of $G_m$. 
	\end{enumerate}
	Equipped with these data, we can write down the Hilbert series 
	\be
	H_{Q(m,\h m,B)}^{T_{m;\h m,B}}(t,\h x)
	\ee
	that counts chiral operators of electric charges $-Q(m,\h m,B)$ in the residual gauge theory, according to the rules of section \ref{sec:HS}. 
	
	The Hilbert series \eqref{HS_3d_N=2} counts dressed monopole operators labelled by their magnetic charges $m$ for the gauge group $G$, which are summed over, and $(\h m;B,0)$ for the global symmetry group $\h G\times G_J\times U(1)_R$, which are held fixed. Putting together all the ingredients discussed so far leads to the {\bf monopole formula} for the Hilbert series \cite{Cremonesi2015a,CremonesiMekareeyaZaffaroni2016}
	\be\label{monopole_formula_N=2}
	H(t,\h x,z;\h m;B)= \sum_{m \in \Gamma_q} t^{R(m,\h m,B)} z^{J(m,\h m,B)} \prod_{\h i} \h x_{\h i}^{\h Q_{\h i}(m,\h m,B)} \cdot H_{Q(m,\h m,B)}^{T_{m;\h m,B}}(t,\h x)~.
	\ee
	The sum is over the quantum lattice of magnetic charges $\Gamma_q$, which is $\Gamma_{G^\vee}/\cW$ or a sublattice thereof if there are nonperturbative effects that lift part of the semiclassical Coulomb branch (see \cite{Cremonesi2015a} and section \ref{sec:3d_Neq2_examples} for details). The powers of the fugacities $t$, $z$ and $\h x$ in the prefactor keep track of the charges \eqref{charges_tot} of a bare monopole operator under the global symmetry, and the Hilbert series $H_{Q(m,\h m,B)}^{T_{m;\h m,B}}$ of the residual gauge theory is the {\bf dressing factor} that keeps track of the charges of the residual matter fields that dress the bare monopole operator. 
	
	We will see some applications of the monopole formula \eqref{monopole_formula_N=2} for the Hilbert series of the moduli space of vacua of three-dimensional $\cN\geq 2$ gauge theories in the following sections.

	\section{Coulomb branch of $3d$ $\cN=4$ gauge theories}\label{sec:CB}
	
	The monopole formula \eqref{monopole_formula_N=2} becomes particularly simple when it is applied to the vector multiplet sector of three-dimensional theories with eight supercharges, \emph{i.e.} $3d$ $\cN=4$ supersymmetry. We will view $3d$ $\cN=4$ gauge theories as special cases of $3d$ $\cN=2$ gauge theories, fixing once and for all an $\cN=2$ subalgebra of the $\cN=4$ supersymmetry algebra. This is equivalent to choosing a particular complex structure out of a $\bP^1$ worth of them. 
	
	Like $4d$ $\cN=2$ theories, $3d$ $\cN=4$ supersymmetric gauge theories are completely specified by the following data:  \begin{enumerate}
		\item {\bf Gauge group}: a compact semisimple Lie group $G$, to which one associates $\cN=4$ vector multiplets $\cV^a$, with $a=1,\dots,\rk(G)$, which contains a gauge connection, three real scalars and fermionic partners. The $\cN=4$ vector multiplet $\cV$ decomposes into an $\cN=2$ vector multiplet $V$ (containing a gauge connection $A$ with curvature $F$, a real scalar $\sigma$ and fermions)
		and an $\cN=2$ chiral multiplet $\Phi$ in the adjoint representation of $G$ (containing a complex scalar that we also call $\Phi$ and fermions).
		\item {\bf Matter content}: a (quaternionic) representation $\cR$ of $G$, to which one associates hypermultiplets $Y^i$, with $i=1,\dots,\dim(\cR)$. The hypermultiplets $Y^i$ decompose into a pair of $\cN=2$ chiral multiplets $X^i$ and $\t X_i$, each containing a complex scalar plus fermions, and transforming in complex conjugate representations $\cR$ and $\bar{\cR}$.%
		\footnote{If the representation $\cR$ is pseudoreal, one can introduce a half-hypermultiplet, which contains half as many degrees of freedom as a standard hypermultiplet. } 
	\end{enumerate}
	The interactions are completely determined by $\cN=4$ supersymmetry. In particular, the superpotential takes the form $W=\t X\Phi X$, where $\Phi$ acts on $X$ in the representation $\cR$ and the projection to the gauge singlet in $\bar{\cR} \otimes \cR$ is implied.
	
	The $\cN=4$ supersymmetry algebra admits an $R$-symmetry automorphism $SU(2)_C\times SU(2)_H$. $SU(2)_C$ acts on vector multiplets, rotating the three real scalars as a triplet.  $SU(2)_H$ acts on hypermultiplets, rotating the two complex scalars $X^i$ and $(\t X_i)^\dagger$ as a doublet. 
	The moduli space of supersymmetric vacua of a $3d$ $\cN=4$ theory is locally of the form $\cM_C\times \cM_H$: on the Coulomb branch $\cM_C$ the scalars in the vector multiplet take expectation value, whereas on the Higgs branch $\cM_H$ the scalars in the hypermultiplets take expectation value. Although Hilbert series methods can be easily applied to the total moduli space of vacua, which includes mixed branches (see for instance \cite{CartaHayashi2016}), we restrict here to the maximal-dimensional components of the Coulomb and Higgs branches. 
	
	Higgs branches of 3d $\cN=4$ supersymmetric gauge theories in three dimensions are identical to the Higgs branches of 4d $\cN=2$ supersymmetric gauge theories with the same gauge group and matter content, being protected against quantum corrections by a non-renormalization theorem \cite{Argyres:1996eh}. In particular they are hyperkähler quotients. Hilbert series of Higgs branches of gauge theories with eight supercharges have been computed in \cite{Hanany:2006uc,Benvenuti2010,Hanany:2011db}.
	
	More interesting results are obtained by applying the logic of this section to Coulomb branches of $3d$ $\cN=4$ gauge theories, which are hyperkähler manifolds of quaternionic dimension $r=\rk(G)$. Hypermultiplet scalars vanish on the Coulomb branch, which is parametrized by monopole operators dressed by scalars $\Phi$ in the vector multiplet of the residual gauge group. Specializing formula \eqref{charges_quant} to the Cartan generator of the $SU(2)_C$ $R$-symmetry, which assigns charge $2$ to $\Phi$ and $0$ to hypermultiplet scalars, one obtains the following $R$-charge for bare monopole operators \cite{Borokhov:2002cg,Gaiotto:2008ak}:
	\be\label{R_monop_Neq4}
	R(m,\h m)= -\sum_{\alpha\in \Delta_+} |\alpha(m)|+\frac{1}{2}\sum_{\rho,\h\rho}|\rho(m)+\h\rho(\h m)| ~.
	\ee
	Taking into account the dressing of monopole operators by the adjoint scalar $\Phi$ in the vector multiplet and the absence of nonperturbative corrections to the superpotential, which is a consequence of $\cN=4$ supersymmetry, the monopole formula \eqref{monopole_formula_N=2} takes the simple form  \cite{Cremonesi:2013lqa,Cremonesi:2014kwa}
	\be\label{monop_form_Neq4}
	H(\tau,z;\h m) = \sum_{m \in \Gamma_G/\cW_G} z^{J(m)} \tau^{2 R(m,\h m)} P_G(\tau^2;m)~,
	\ee
	where the fugacity $\tau=t^{1/2}$ is introduced to have integer powers, and the dressing factor
	\be\label{dressing}
	P_G(t;m)=\prod_{i=1}^r \frac{1}{1-t^{d_i(G_m)}}
	\ee
	counts polynomials in the Casimir invariants of the residual gauge group $G_m$, as in formula \eqref{HS_adjoint} with $G$ replaced by $G_m$. Note that  the background magnetic charge $B$ for the topological symmetry was set to zero in \eqref{monop_form_Neq4}, otherwise there are no gauge invariant monopole operators dressed by $\Phi$ only; this agrees with the well known fact that a Fayet-Iliopoulos parameter lifts the Coulomb branch.
	
	It should be noted that the monopole formula \eqref{monop_form_Neq4} yields a well defined Taylor series in $\tau$ provided the theory is \emph{good} or \emph{ugly} in the terminology of \cite{Gaiotto:2008ak}.%
	\footnote{For \emph{bad} theories \eqref{monop_form_Neq4} is not a convergent Taylor series, because there are infinitely many monopole operators with the same charges. This problem can be bypassed by adding extra hypermultiplets to make the theory good or ugly, with a large background magnetic charge for the flavour symmetry that acts on them only. The background charge ensures that the common eigenspaces of the Cartan subalgebra of the global symmetry have finite dimension and  serves at the same time as a cut-off.} 
	Then the conformal dimension of chiral gauge invariant operators in the infrared superconformal field theory is equal to their $R$-charge. 
	
	Recently, a mathematical interpretation of the monopole formula \eqref{monop_form_Neq4} for the Hilbert series of the Coulomb branch of 3d $\cN=4$ gauge theories has been provided by Nakajima \cite{Nakajima2015a}, leading to a number of interesting mathematical developments, for which I refer to Nakajima's talk at this conference.
	
	A few additional remarks on \eqref{monop_form_Neq4} are in order. First of all, $H(t,z;0)$ is the Hilbert series that counts holomorphic function on the Coulomb branch of the low energy superconformal field theory, which is a cone. Conversely, $\h m\neq 0$ corresponds to turning on real masses that resolve the singularity. 
	
	Secondly, it has been shown in \cite{Razamat:2014pta} that the Hilbert series of the Coulomb branch given by  \eqref{monop_form_Neq4} can also be obtained as a particular (and more easily computable) limit of the superconformal index. This means that in this limit the index only receives contribution from the scalar chiral operators which are counted by the Hilbert series, and not by other protected operators.
	
	Thirdly, as we will see in some of the examples below, one can often deduce the charges of generators and relations of the Coulomb branch chiral ring by resumming \eqref{monop_form_Neq4} and applying plethystic techniques. In this respect the Hilbert series provides complementary information to the more recent physical description of the Coulomb branch in \cite{BullimoreDimofteGaiotto2015}: the latter allows to construct the relations explicitly, but in practice determining the generators of the chiral ring and of the ideal of relations can be difficult if the Coulomb branch is not a complete intersection. Instead the Hilbert series can be computed as easily for complete as for non-complete intersections.
	
	Finally, the Hilbert series \eqref{monop_form_Neq4} is sensitive to resolutions of the singularity (K\"ahler deformations) through the dependence on $\h m$ and $B$, but is insensitive to complex structure deformations. On the other hand the formalism of \cite{BullimoreDimofteGaiotto2015} is sensitive to complex structure deformations but insensitive to resolutions.
	
	\subsection{Examples}\label{subsec:examples_3dNeq4}
	We conclude this section with a few examples of Hilbert series of $3d$ $\cN=4$ theories with interesting Coulomb branches,  restricting for simplicity to zero background magnetic charges. Examples with $\h m\neq 0$ can be found in \cite{Cremonesi:2014kwa,Cremonesi:2014uva}. 
	
	\begin{figure}
		\centering
		\begin{tikzpicture}[baseline, baseline,font=\footnotesize, scale=1]
		\begin{scope}[auto,%
		every node/.style={draw, minimum size=.7 cm}, node distance=1.5cm];
		\node[circle] (U1) at (0, 0) {$k$};
		\node[rectangle] (FQ) at (2,0) {$N$};
		\end{scope}
		\draw[draw=black,solid,line width=0.2mm,-]  (U1) to  (FQ) ; 
		\end{tikzpicture}
		\caption{8-supercharge quiver diagram for $U(k)$ SQCD with $N$ flavours. \label{fig:SQCD_Neq4_quiver}}
	\end{figure}
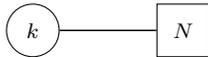

	It is well known that the Coulomb branch of $3d$ $\cN=4$ SQED, with $G=U(1)$ and $N$ flavours of hypermultiplets of charge $1$, is the $A_{N-1}$ singularity $\bC^2/\bZ_N$ \cite{Intriligator:1996ex}. This result is easily recovered by computing the Hilbert series \cite{Cremonesi:2013lqa}
	\be\label{SQED}
	H(\tau,z)= \frac{1}{1-\tau^2} \sum_{m \in \bZ} z^m \tau^{N|m|}=\PE[\tau^2 + (z+z^{-1})\tau^N - \tau^{2N}]~,
	\ee
	which shows that the Coulomb branch chiral ring is generated by three operators $\Phi$, $V_+\equiv V_1$ and $V_-\equiv V_{-1}$, subject to a single relation $V_+ V_-=\Phi^N$. This is the well-known algebraic description of the $A_{N-1}$ singularity.
	
	The computation is easily generalized to SQCD theories with $G=U(k)$ and $N\geq 2k-1$ flavours of fundamental hypermultiplets, which are summarized by ``quiver diagram'' in figure \ref{fig:SQCD_Neq4_quiver}, which uses the the eight-supercharge notation where a circular node denotes a unitary gauge group, a square node denotes a unitary flavour group, and edges denote hypermultiplets in the bifundamental representation. The Hilbert series of the Coulomb branch is \cite{Cremonesi:2013lqa}
	\be\label{SQCD}
	\begin{split}
		H(\tau,z)&= \sum_{m_1\geq\dots\geq m_k} P_{U(k)}(\tau^2;m) z^{\sum_i m_i}~ \tau^{-2\sum_{i<j}(m_i-m_j)+N \sum_i |m_i|}=\\
		&= \PE\Big[ \sum_{j=1}^k \big( \tau^{2j} + (z+z^{-1}) \tau^{N+2(j-k)}-\tau^{2(N+j-k)}  \big)\Big]	~.
	\end{split}
	\ee
	The dressing factor $P_{U(k)}$ can be written in terms of a partition of $k$ that encodes how many magnetic charges $m_i$ are equal, see appendix A of \cite{Cremonesi:2013lqa} for the explicit form. The final expression in (\ref{SQCD}) shows that the Coulomb branch is a complete intersection: there are $3k$ generators ($k$ Casimirs and $k+k$ dressed monopole operators of topological charges $\pm 1$) subject to $k$ relations, whose explicit form was later obtained in \cite{BullimoreDimofteGaiotto2015}.
	
	\begin{figure}
		\centering
		\begin{tikzpicture}[baseline, baseline,font=\footnotesize, scale=1.3]
		\begin{scope}[auto,%
		every node/.style={draw, minimum size=.7 cm}, node distance=1.cm];
		\node[rectangle] (-1) at (-1, 0) {$1$};
		\node[circle] (0) at (0,0) {$k$};
		\node[circle] (1) at (1,0) {2$k$};
		\node[circle] (2) at (2,0) {$k$};
		\end{scope}
		\draw[draw=black,solid,line width=0.2mm,-]  (-1) to (0) ; 
		\draw[draw=black,solid,line width=0.2mm,-]  (0) to (1) ; 
		\draw[draw=black,solid,line width=0.2mm,-]  (1) to (2) ; 
		\draw[draw=black,solid,line width=0.2mm,-]  (1.25,0.1) to (1.75,0.1) ; 
		\draw[draw=black,solid,line width=0.2mm,-]  (1.25,-0.1) to (1.75,-0.1);				
		\draw[draw=black,solid,line width=0.2mm,-]  (1.4,0.2) to (1.6,0) ; 
		\draw[draw=black,solid,line width=0.2mm,-]  (1.4,-0.2) to (1.6,0);	\
		\node at (-1, -.5) {$-1$};
		\node at (0, -.5) {$0$};
		\node at (1, -.5) {$1$};
		\node at (2, -.5) {$2$};
		\end{tikzpicture}
		\caption{Generalized affine Dynkin quiver for $k$  $G_2$ instantons.}
		\label{fig:G2}
	\end{figure}
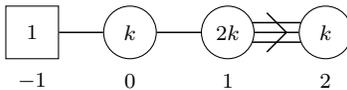

	A very interesting application is to theories whose Coulomb branches are moduli spaces $\cM_{\cG,k}$ of $k$ $\cG$-instantons on $\bC^2$ \cite{Cremonesi:2014xha}. These theories are described by (generalized) quiver diagrams which are affine Dynkin diagrams of $\cG$, with a $U(1)$ flavour node attached to the zeroth root. See figure \ref{fig:G2} for the example of $\cG=G_2$.
	If $\cG$ has a non-simply-laced Lie algebra, the field theory has no known Lagrangian description, but a monopole formula for the Hilbert series of their Coulomb branches was conjectured in \cite{Cremonesi:2014xha} based on brane constructions and the action of outer automorphisms of simply laced Lie algebras. The proposal involves a minimal modification of \eqref{monop_form_Neq4} that deals with multiple bonds in the Dynkin diagrams. For instance, the triple bond in the affine Dynkin diagram of $G_2$ leads to the contribution
	\be\label{multi_bond}
	\Delta R(m)= -\frac{1}{2}\sum_{i=1}^{2k} \sum_{j=1}^{k} |3m^{(1)}_i- m^{(2)}_j|
	\ee
	to the $R$-charge of monopole operators, where the factor of $3$ accounts for the triple bond directed from node $1$ to node $2$. 
	
	This conjectured modification ensures that the affine Dynkin quiver without the flavour node is balanced, which is crucial for the enhancement of the topological symmetry to a nonabelian gauge group $\cG\times SU(2)$, and passes several other nontrivial consistency checks.
	Most importantly, it provides a uniform description of instanton moduli spaces which is purely based on group theoretic data and is alternative to the ADHM  construction \cite{Atiyah:1978ri} that applies to classical gauge groups $\cG$. See \cite{BravermanFinkelbergNakajima2016a} for a further mathematical exploration of this modified monopole formula.
	
	Using this general description, it was shown in \cite{Cremonesi:2014xha} that the Hilbert series of the moduli space $\cM_{\cG,k}$ of $k$ $\cG$-instantons has the following perturbative expansion, that holds for any $\cG$: 
	\be\label{instantons}
	H_{\cM_{\cG,k}}(t,x,y)=\PE\bigg[ \sum_{p=0}^{k-1} \big([p+1;0]_{x,y}t^{p+1}+ [p;\mathrm{Ad}]_{x,y} t^{p+2} \big)- \dots t^{k+2}+\dots  \bigg]~.
	\ee
	Here $x$ is an $SU(2)$ fugacity, $y$ collectively denotes $\cG$ fugacities, and $\mathrm{Ad}$ denotes the adjoint representation of $\cG$. The first $2k$ positive terms in the argument of the plethystic exponential are generators. In particular $[1;0]_{x,y}t$ corresponds to the centre of mass of the $k$  instantons, and $([2;0]_{x,y}+ [0;\mathrm{Ad}]_{x,y}) t^{2}$ correspond to the moment maps of the $SU(2)\times \cG$ symmetry. The first relation appears at order $t^{k+2}$. It would be interesting to see whether this uniform analysis of instanton moduli spaces can be pushed further to fully determine generators and relations.

	\section{Moduli spaces of $3d$ $\cN=2$ theories and Hilbert series}\label{sec:3d_Neq2_examples}
	
	The rationale for counting dressed monopole operators presented in section \ref{sec:HS_3d} was initially developed to study Coulomb branches of $3d$ $\cN=4$ theories, but it was soon realized that it applies just as well to the moduli space of vacua of theories with $\cN\geq 2$ supersymmetry. In this section we present a few examples involving Maxwell, Yang-Mills \cite{Cremonesi2015a}
	and Chern-Simons theories \cite{CremonesiMekareeyaZaffaroni2016}. 
	
	\subsection{Maxwell and Yang-Mills theories}
	
	We begin with CP-invariant theories, with no Chern-Simons interactions. For simplicity we set all background magnetic charges $\h m$ and $B$ to zero and take the superpotential to vanish (see \cite{CremonesiMekareeyaZaffaroni2016} for the generalization).
	Then the Hilbert series that counts gauge invariant chiral operators, or equivalently holomorphic functions on the moduli space of supersymmetric vacua
	$\cM$, takes the general form \cite{Cremonesi2015a,Hanany:2015via}
	\be\label{YangMills}
	\begin{split}
		H(t,z,\h x) &= \sum_{m \in \Gamma_q} z^{J(m)} \bigg(\prod_{i=1}^r \oint \frac{dx_i}{2\pi i x_i}\bigg) \prod_{\alpha \in \Delta_+} (1-x^\alpha)^{\delta_{\alpha(m),0}} t^{-|\alpha(m)|} \cdot\\
		& \qquad\quad \cdot \prod_{\rho,\h\rho} \left(t^{r_{\rho,\h\rho}-1} x^\rho \h x^{\h \rho} \right)^{-\frac{1}{2}|\rho(m)|} \PE\big[\delta_{\rho(m),0} t^{r_{\rho,\h\rho}} x^\rho \h x^{\h \rho} \big]~,
	\end{split}
	\ee
	where the product over $\rho$ and $\h\rho$ runs over all weights of the representation $\cR\times \h\cR$ of the matter fields under the gauge and flavour symmetry, and $r_{\rho,\h\rho}$ are the $R$-charges of the matter fields, which are constant in each irreducible representation. As anticipated below \eqref{monopole_formula_N=2}, due to nonperturbative effects that lift part of the classical Coulomb branch \cite{BoerHoriOz1997}, the sum over magnetic charges is restricted to a sublattice of $\Gamma_{G^\vee}/\cW$, the quantum lattice of magnetic charges $\Gamma_q$, which in this case is 
	\be\label{Gamma_q}
	\Gamma_q = \{ m \in \Gamma_{G^\vee}/\cW ~ |~ \sum_{\rho} \rho(\alpha_i^\vee) \mathrm{sign}\rho(m)\neq 0 ~\vee~ \alpha_i(m)=0 ~~ \forall i=1,\dots,r  \}~,
	\ee
	where $\alpha_i$ and $\alpha_i^\vee$ are simple roots and coroots of the gauge group $G$.
	
	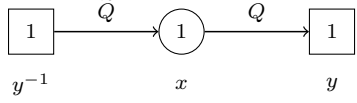
\begin{figure}
		\begin{tikzpicture}[baseline, baseline,font=\footnotesize, scale=1]
		\begin{scope}[auto,%
		every node/.style={draw, minimum size=.6 cm}, node distance=1.5cm];
		\node[rectangle] (FQt) at (-2,0) {$1$};
		\node[circle] (U1) at (0, 0) {$1$};
		\node[rectangle] (FQ) at (2,0) {$1$};
		\end{scope}
		\draw[draw=black,solid,line width=0.2mm,->]  (U1) to node[midway,above] {$Q$}node[midway,above] {}  (FQ) ; 
		\draw[draw=black,solid,line width=0.2mm,->]  (FQt) to node[midway,above] {$\t Q$}node[midway,above] {}  (U1) ;
		
		\node at (-2,-0.7) {$y^{-1}$};
		\node at (0,-0.7) {$x$};
		\node at (2,-0.7) {$y$};
		
		\end{tikzpicture}
		\caption{Quiver diagram and fugacities for SQED with one flavour. \label{SQED_1}}
	\end{figure}
	
	\begin{figure}
		\begin{tikzpicture}[baseline, baseline,font=\footnotesize, scale=1]
		\begin{scope}[auto,%
		every node/.style={draw, minimum size=.6 cm}, node distance=1.5cm];
		\node[rectangle] (FQt) at (-2,0) {$N_f$};
		\node[circle] (U1) at (0, 0) {$N$};
		\node[rectangle] (FQ) at (2,0) {$N_f$};
		\end{scope}
		\draw[draw=black,solid,line width=0.2mm,->]  (U1) to node[midway,above] {$Q$}node[midway,above] {}  (FQ) ; 
		\draw[draw=black,solid,line width=0.2mm,->]  (FQt) to node[midway,above] {$\t Q$}node[midway,above] {}  (U1) ;
		
		\end{tikzpicture}
		\caption{Quiver diagram for $U(N)$ SQCD with $N_f$ flavours. \label{SQCD_2}}
	\end{figure}
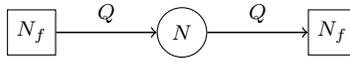

	For instance, applying formula \eqref{YangMills} to $3d$ $\cN=2$ SQED with one flavour, whose quiver diagram is in figure \ref{SQED_1}, one obtains the Hilbert series 
	\be\label{Neq2_SQED_1}
	\begin{split}
		H(t,z,y)&=\sum_{m\in\bZ} z^m t^{(1-r)|m|}y^{-|m|} \oint \frac{dx}{2\pi i x} \PE\left[ \delta_{m,0} t^r y(x+\frac{1}{x})  \right]=\\
		&= \frac{1}{1-y^2 t^{2r}}+\frac{1}{1-z y^{-1}t^{1-r}}+\frac{1}{1-z^{-1} y^{-1}t^{1-r}}-2~,
	\end{split}
	\ee
	where $y$ is a fugacity for the axial $U(1)$ symmetry under which the matter fields $Q$ and $\t Q$ have charge $1$, and equal $R$-charge $r$ has been assigned to $Q$ and $\t Q$. The final result shows that the moduli space consists of three-one dimensional components meeting at a point, generated by the meson $M=\tilde Q Q$ and the monopole operators $V_\pm \equiv V_{\pm 1}$, subject to the relations $V_+ V_- = V_+ M = V_- M=0$. This reproduces the moduli space and chiral ring of the $XYZ$ model \eqref{XYZ}, of which SQED with one flavour is known to be dual \cite{AharonyHananyIntriligatorEtAl1997,BoerHoriOz1997}.
	
	Even more instructive is the case of $U(N)$ SQED with $N_f$ flavours of fundamental and antifundamental matter, see figure \ref{SQCD_2}. The nonabelian gauge dynamics induces nonperturbative corrections to the superpotential, which make the quantum lattice \eqref{Gamma_q} two-dimensional:
	\be\label{Gamma_q_SQCD}
	\Gamma_q = \{ m=(m_1,0,\dots,0,m_N) \in \bZ^N ~ |~  m_1\geq 0 \geq m_N \}~.
	\ee
	The Hilbert series can be computed straightforwardly \cite{Cremonesi2015a}, but we do not report the result here. Suffices to say that the chiral ring of the theory follows immediately from the structure of the residual gauge theories associated to lattice points in \eqref{Gamma_q_SQCD} and the knowledge of their moduli spaces. Over the point $m_1=m_N=0$, the residual theory is nothing but $U(N)$ SQCD with $N_f$ flavours; over the 1-dimensional boundary the residual theory is $U(N-1)$ SQCD with $N_f$ flavours (plus a free $U(1)$ theory); and over the two-dimensional interior the residual theory is $U(N-2)$ SQCD with $N_f$ flavours (plus two free $U(1)$ theories). It follows almost immediately that the chiral ring of the theory is generated by the $N_f\times N_f$ meson matrix $M=\tilde Q Q$ and the two monopole operators $V_+\equiv V_{(1,0,\dots,0)}$ and $V_-\equiv V_{(0,\dots,0,-1)}$, subject to the chiral ring relations
	\be\label{SQCD3_chiral_ring}
	\mathrm{minor}_{N+1}(M)=0~, \quad V_\pm \mathrm{minor}_{N}(M)=0~, \quad V_+ V_- \mathrm{minor}_{N-1}(M)=0~  
	\ee
	which constrain the rank of the meson matrix and monopole operators. This result is confirmed by explicit computation of the Hilbert series.
	
	This structure of the moduli space of vacua was argued earlier in \cite{Aharony1997} by invoking an extra nonperturbative superpotential involving mesons and monopole operators, which however is singular at certain subloci of the moduli space. We learn from this example that the moduli spaces and the chiral rings of the nonabelian $\cN=2$ Yang-Mills theories can be successfully analysed by looking at dressed monopole operators and using well-defined nonperturbative superpotentials that partially lift the Coulomb branch, without resorting to singular nonperturbative superpotentials.

	\subsection{Chern-Simons theories}
	
	The formalism presented in section \ref{sec:HS_3d} can also be applied to CP-violating gauge theories with Chern-Simons interactions \eqref{CS_susy}, which were studied intensively following \cite{Aharony:2008ug} since they often appear on the worldvolume of M2-branes in M-theory.
	
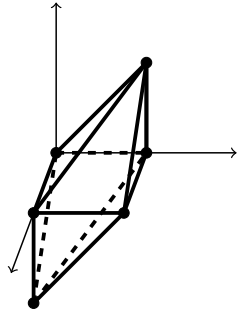
\begin{figure}[t]
		\centering 
		\begin{tikzpicture}
		[baseline, font=\footnotesize, scale=1]
		
		\filldraw[black] (0,0) circle (2pt) ;
		\filldraw[black] (1.2,0) circle (2pt);
		\filldraw[black] (-.3,-.8) circle (2pt) ;
		\filldraw[black] (.9,-.8) circle (2pt) ;
		\filldraw[black] (1.2,1.2) circle (2pt);
		\filldraw[black] (-.3,-2) circle (2pt) ;

		\draw[draw=black,dashed,line width=0.5mm,-]  (0,0) to (1.2,0) ;  
		\draw[draw=black,solid,line width=0.5mm,-]  (0,0) to (-.3,-.8) ;  
		\draw[draw=black,solid,line width=0.5mm,-]  (1.2,0) to (.9,-.8) ;  
		\draw[draw=black,solid,line width=0.5mm,-]  (-.3,-.8) to (.9,-.8) ;  
		\draw[draw=black,solid,line width=0.5mm,-]  (1.2,1.2) to (1.2,0) ;  
		\draw[draw=black,solid,line width=0.5mm,-]  (1.2,1.2) to (0,0) ;  
		\draw[draw=black,solid,line width=0.5mm,-]  (1.2,1.2) to (1.2,0) ;  
		\draw[draw=black,solid,line width=0.5mm,-]  (1.2,1.2) to (-.3,-.8) ;  
		\draw[draw=black,solid,line width=0.5mm,-]  (1.2,1.2) to (.9,-.8) ;  
		\draw[draw=black,dashed,line width=0.5mm,-]  (-.3,-2) to (0,0) ;  
		\draw[draw=black,dashed,line width=0.5mm,-]  (-.3,-2) to (1.2,0) ;  
		\draw[draw=black,solid,line width=0.5mm,-]  (-.3,-2) to (-.3,-.8) ;  
		\draw[draw=black,solid,line width=0.5mm,-]  (-.3,-2) to (.9,-.8) ;

		\draw[draw=black,solid,line width=0.2mm,->]  (0,0) to (-.6,-1.6) ;  
		\draw[draw=black,solid,line width=0.2mm,->]  (0,0) to (2.4,0) ;  
		\draw[draw=black,solid,line width=0.2mm,->]  (0,0) to (0,2) ;   		
		\end{tikzpicture}
		\caption{Toric diagram of $C(Q^{1,1,1})$. 	\label{fig:toric_Q111}}
	\end{figure}

	As an example, we consider here the $3d$ $\cN=2$ gauge theory on the worldvolume of M2-branes probing the Calabi-Yau cone over the toric Sasaki-Einstein $7$-fold $Q^{1,1,1}$, which is a circle fibration over $\bP^1\times\bP^1\times\bP^1$. The toric diagram of the Calabi-Yau Gorenstein singularity $C(Q^{1,1,1})$ is a convex polytope in three dimensions, which is shown in figure \ref{fig:toric_Q111}.

	\begin{figure}[t]
		\centering 
		\begin{tikzpicture}[baseline, font=\footnotesize, scale=1]
		\begin{scope}[auto,%
		every node/.style={draw, minimum size=.6cm}, node distance=2cm];
		\node[circle] (USp2k) at (-0.1, 0) {$N_0$};
		\node[circle, right=of USp2k] (BN)  {$N_0$};
		\node[rectangle] (F1) at (1.3, 1.5) {$1$};
		\node[rectangle] (F2) at (1.3, 2.2) {$1$};
		\end{scope}
		\draw[draw=black,solid,line width=0.2mm,->>]  (USp2k) to[bend right=30] node[midway,above] {$A_{1,2}$}node[midway,above] {}  (BN) ;  
		\draw[draw=black,solid,line width=0.2mm,<<-]  (USp2k) to[bend left=30] node[midway,below] {$B_{1,2}$} node[midway,above] {} (BN) ; 
		\draw[draw=black,solid,line width=0.2mm,->]  (F1) to[bend left=20] node[midway,left] {$p_1$}node[midway,above] {}  (BN) ;   
		\draw[draw=black,solid,line width=0.2mm,->]  (F2) to[bend left=20] node[midway,right] {$p_2$}node[midway,above] {}  (BN) ;   
		\draw[draw=black,solid,line width=0.2mm,->]  (USp2k) to[bend left=20] node[midway,right] {$q_1$}node[midway,above] {}  (F1) ;   
		\draw[draw=black,solid,line width=0.2mm,->]  (USp2k) to[bend left=20] node[midway,left] {$q_2$}node[midway,above] {}  (F2) ;   
		\end{tikzpicture}
		\caption{Quiver diagram of the worldvolume theory on $N$ M2-branes probing $C(Q^{1,1,1})$. The subscripts denote bare CS levels.
			\label{fig:quiver_Q111}}
	\end{figure}
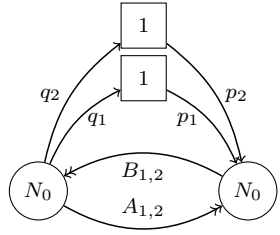


	The gauge theory on the worldvolume of M2-branes is specified by the quiver diagram in figure \ref{fig:quiver_Q111} and by the superpotential \cite{Benini2010,Jafferis:2009th}
	\be\label{W_Q111}
	W = \mathrm{Tr}(A_1 B_1 A_2 B_2 - A_1 B_2 A_2 B_1) + p_1 B_1 q_1 + p_2 B_2 q_2~.
	\ee
	The bare Chern-Simons levels for the two gauge groups in the Lagrangian vanish, $k_1=k_2=0$, but effective Chern-Simons levels \eqref{eff_CS} are radiatively generated once the fundamental flavours $p_i$ and $q_i$ gain a real mass and are integrated out.

	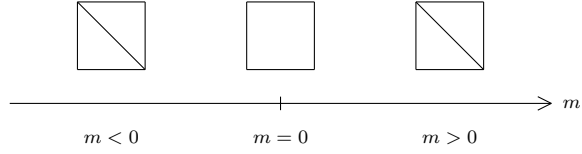
\begin{figure}
		\begin{tikzpicture}[baseline, baseline,font=\footnotesize, scale=0.9, transform shape]
		\draw (-3,1.5)--(-2,1.5); 
		\draw (-3,.5)--(-2,.5);
		\draw (-3,1.5)--(-3,.5); 
		\draw (-2,1.5)--(-2,.5);
		\draw (-3,1.5)--(-2,.5);
		
		\draw (-.5,1.5)--(.5,1.5); 
		\draw (-.5,.5)--(.5,.5);
		\draw (-.5,1.5)--(-.5,.5); 
		\draw (.5,1.5)--(.5,.5);
		
		\draw (3,1.5)--(2,1.5); 
		\draw (3,.5)--(2,.5);
		\draw (3,1.5)--(3,.5); 
		\draw (2,1.5)--(2,.5);
		\draw (2,1.5)--(3,.5);
		
		\draw (-4,0)--(4,0);
		\draw (3.8,.1)--(4,0);
		\draw (3.8,-.1)--(4,0);	
		\draw (0,.1)--(0,-.1);
		
		\node at (4.3,0) {$m$};
		
		\node at (-2.5,-0.5) {$m<0$};
		\node at (0,-0.5) {$m=0$};
		\node at (2.5,-0.5) {$m>0$};
		
		\end{tikzpicture}
		\caption{Line bundles and resolutions of the conifold from holomorphic functions on $C(Q^{1,1,1})$. The toric diagrams of the resolved conifolds can be obtained by projecting the toric diagram of $C(Q^{1,1,1})$ of figure \ref{fig:toric_Q111} in the vertical direction. \label{fig_resol_conif_from_Q111}	}	
	\end{figure}
	
	The Hilbert series of the moduli space of vacua is easily computed for the abelian theory on $N=1$ M2-brane. Setting the background magnetic charges to zero and the dynamical magnetic charges equal $m_1=m_2=m$ (there are no gauge invariant dressed monopole operators if $m_1\neq m_2$), the Hilbert series of the gauge theory \eqref{monopole_formula_N=2} becomes \cite{CremonesiMekareeyaZaffaroni2016}
	\be\label{HS_Q111}
	\begin{split}
		H(t,u,v,z) &= \sum_{m \in \bZ} z^m t^{\frac{1}{2}|m|} 
		\oint \frac{dx}{2\pi i x} x^{-|m|} 
		\PE\left[ t^{\frac{1}{2}}x\left(u+\frac{1}{u}\right) + t^{\frac{1}{2}}\frac{1}{x}\left(v+\frac{1}{v}\right) \right]\\
		&= \sum_{m\in\bZ} z^m t^{\frac{1}{2}|m|} g_1(t^{1/2},u,v;|m|) = \sum_{n=0}^\infty [n;n;n]_{\alpha,\beta,\gamma} t^n
	\end{split}
	\ee
	where we have used \eqref{resolved_conifold} in the second line. The final result is nothing but the Hilbert series of the cone over $Q^{1,1,1}$, which can be computed alternatively using the toric description of the geometry: the moduli space of the gauge theory on the M2-branes is the transverse space that they probe. The integral in the first line shows that the residual theory when $m_1=m_2=m$ is the worldvolume theory for a D-brane on the conifold \cite{Klebanov:1998hh}, which up to a decoupled $U(1)$ is the SQED theory with $2$ flavours discussed at the end of section \ref{sec:HS}. The conifold arises here as the K\"ahler quotient of the cone over $Q^{1,1,1}$ by the $U(1)$ action that corresponds to the vertical direction in the toric diagram \ref{fig:toric_Q111}. Going from the first to the second line we have used the baryonic Hilbert series \eqref{resolved_conifold} that counts holomorphic sections of line bundles on the conifold, with baryonic charge $B=|m|$, due to the effective Chern-Simons levels $k_1^{\rm eff}=-k_2^{\rm eff}=\mathrm{sign}(m)$. The last equality relates the sum over holomorphic sections of line bundles $\oplus_m \cO(|m|D)$ on the conifold to holomorphic functions on the cone $C(Q^{1,1,1})$ (see figure \ref{fig_resol_conif_from_Q111} and compare with figure \ref{fig_resol_conif}). 
	The final expression in \eqref{HS_Q111}, involving $SU(2)^3$ characters in terms of the fugacities $\alpha=u$, $\beta=(v/z)^{1/2} $ and $\beta=(vz)^{1/2}$, shows a global symmetry enhancement to $SU(2)^3\times U(1)_R$ and reproduces the Hilbert series of the cone over $Q^{1,1,1}$. Generators and relations of the ring can be easily extracted from the final formula \cite{CremonesiMekareeyaZaffaroni2016}.
	
	The computation \eqref{HS_Q111} can be easily generalized to include non-zero background magnetic charges: the Hilbert series then counts holomorphic section of toric line bundles on $C(Q^{1,1,1})$ \cite{CremonesiMekareeyaZaffaroni2016}, which are in correspondence with resolutions of the cone \cite{Benini2011}. It can also be extended to the case of $N>1$ M2-branes, in which case one can show by means of the Hilbert series that the moduli space is $\cM=\mathrm{Sym}^N C(Q^{1,1,1})$, as expected from string theory considerations \cite{CremonesiMekareeyaZaffaroni2016}. 
	
	Many more examples of M2-brane theories with $\cN=2$ and $\cN=3$ supersymmetries have been studied using the Hilbert series formalism in \cite{CremonesiMekareeyaZaffaroni2016}. Crucially, the Hilbert series can be computed by counting gauge invariant dressed monopole operators, and information on the chiral ring can be extracted from it, with no need to assume the form of the quantum relation between monopole operators as was originally done in \cite{Benini2010,Jafferis:2009th}.

	\section{Conclusion}\label{sec:conclusion}
	
	I have presented here a general formalism to count gauge invariant chiral operators that parametrize moduli spaces of supersymmetric vacua of three-dimensional $\cN\geq 2$ gauge theories. The formalism enumerates gauge invariant dressed monopole operators and packages the information in a generating function called the Hilbert series. By applying plethystic techniques to the Hilbert series, one can extract the charges of the generators of the chiral ring and of the ideal of relations. The formalism simplifies considerably the study of the chiral ring of three-dimensional supersymmetric gauge theories: in favourable cases it determines it completely, and more generally it reduces the problem to determining a finite number of coefficients.
	
	A number of interesting open questions are raised by these results:
	\begin{enumerate}
		\item For the Coulomb branch of $3d$ $\cN=4$ gauge theories, a procedure to determine the chiral ring relations, including all coefficients, has been put forward by Bullimore, Dimofte and Gaiotto \cite{BullimoreDimofteGaiotto2015}. Can that construction be generalized to theories with lower supersymmetry, for which I have reviewed how to compute the Hilbert series here?
		\item In a parallel development, a mathematical definition of the Coulomb branch of $3d$ $\cN=4$ gauge theories has been introduced by Nakajima and collaborators \cite{Nakajima2015a,BravermanFinkelbergNakajima2016,BravermanFinkelbergNakajima2016a}, building on the monopole formula \eqref{monop_form_Neq4} for the Hilbert series. Can one similarly define moduli spaces of vacua of $3d$ $\cN\geq 2$ gauge theories mathematically, building on formula \eqref{monopole_formula_N=2}?
		\item A path integral definition of the Hilbert series of the Coulomb branch as an index has been provided for $3d$ $\cN=4$ gauge theories \cite{Razamat:2014pta,ClossetKim2016}. Can the Hilbert series of the moduli space of vacua of $3d$ $\cN\geq 2$ theory be defined as an index, at least when the $F$-flat moduli space $\cF$ is a complete intersection?
	\end{enumerate} 
	I hope that some of these questions can be answered in the near future.

	\section*{Acknowledgements}

	It is a pleasure to thank Giulia Ferlito, Amihay Hanany, Noppadol Mekareeya and Alberto Zaffaroni for a fruitful and enjoyable collaboration on most of the work reported here, and the organizers of String--Math 2016 for giving me the opportunity to present these results in a highly stimulating environment. I would also like to thank the referee for pointing out several typos in the original version.

	\bibliographystyle{amsplain}
	\bibliography{ref}
	
\end{document}